\documentclass[aps,showpacs,twocolumn,prl]{revtex4-1}
\usepackage{psfig}
\psfigurepath{./}
\usepackage[utf8]{inputenc} 

\newcommand{\Fig}[1]{Fig.~\ref{#1}}     
\newcommand{\Figs}[1]{Figs.~\ref{#1}}     
\newcommand{\lW}{\linewidth}
\begin{document}
\title{Comment on ``Penrose Tilings as Jammed Solids''}
\author{Cristian F.~Moukarzel~\footnote{email address:
    cristian@mda.cinvestav.mx}} \author{Gerardo G.~Naumis}
\affiliation{Depto.~de Física Aplicada, CINVESTAV del IPN, 97310
  M\'erida, Yucat\'an, M\'exico, } 
\affiliation{Depto.~de Física-Química, Instituto de Física, UNAM, AP
  20-364, 01000, México DF, Mexico.}
\maketitle
\begin{figure}
  \centerline{
    \psfig{file=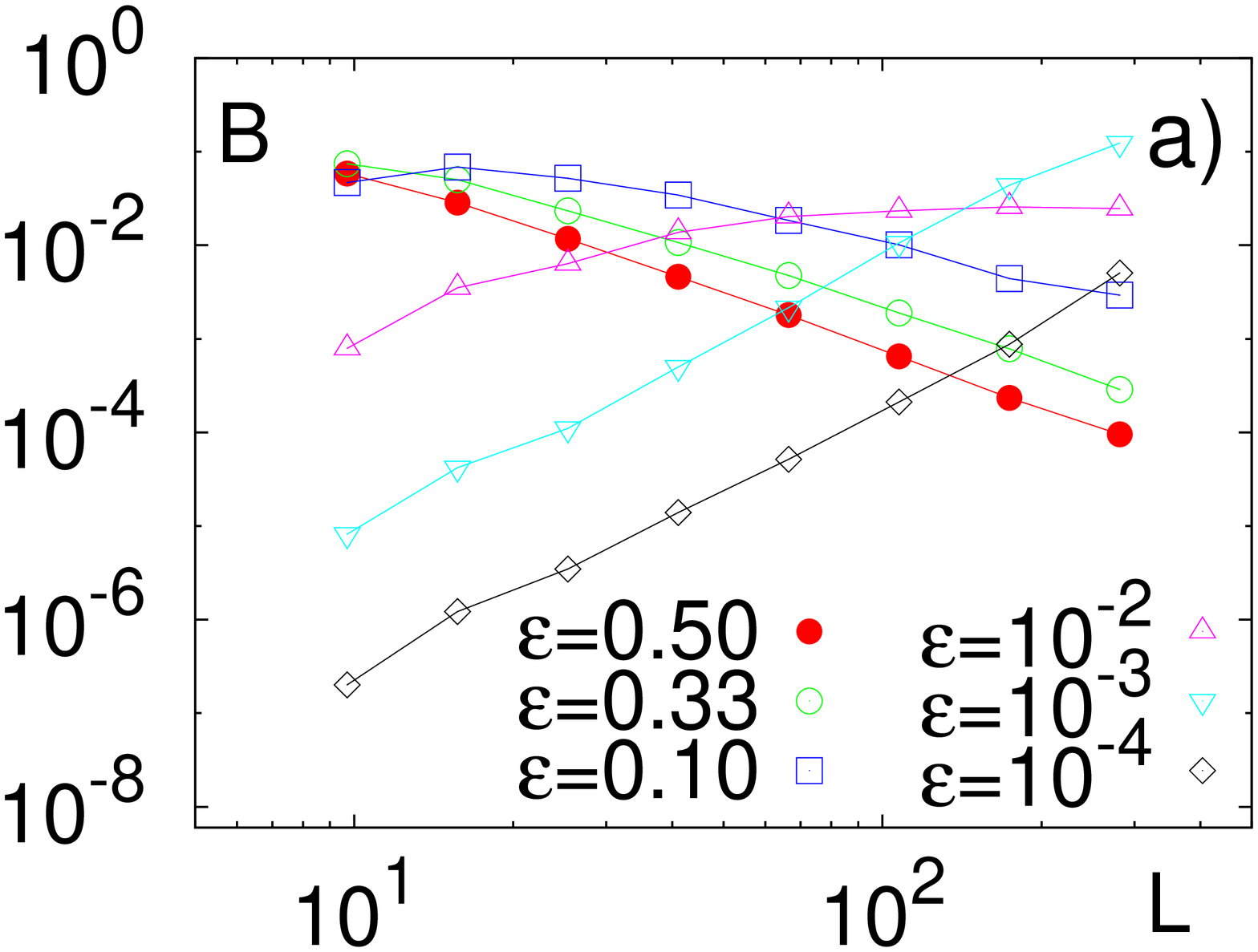,angle=270,width=0.5\lW}
    \psfig{file=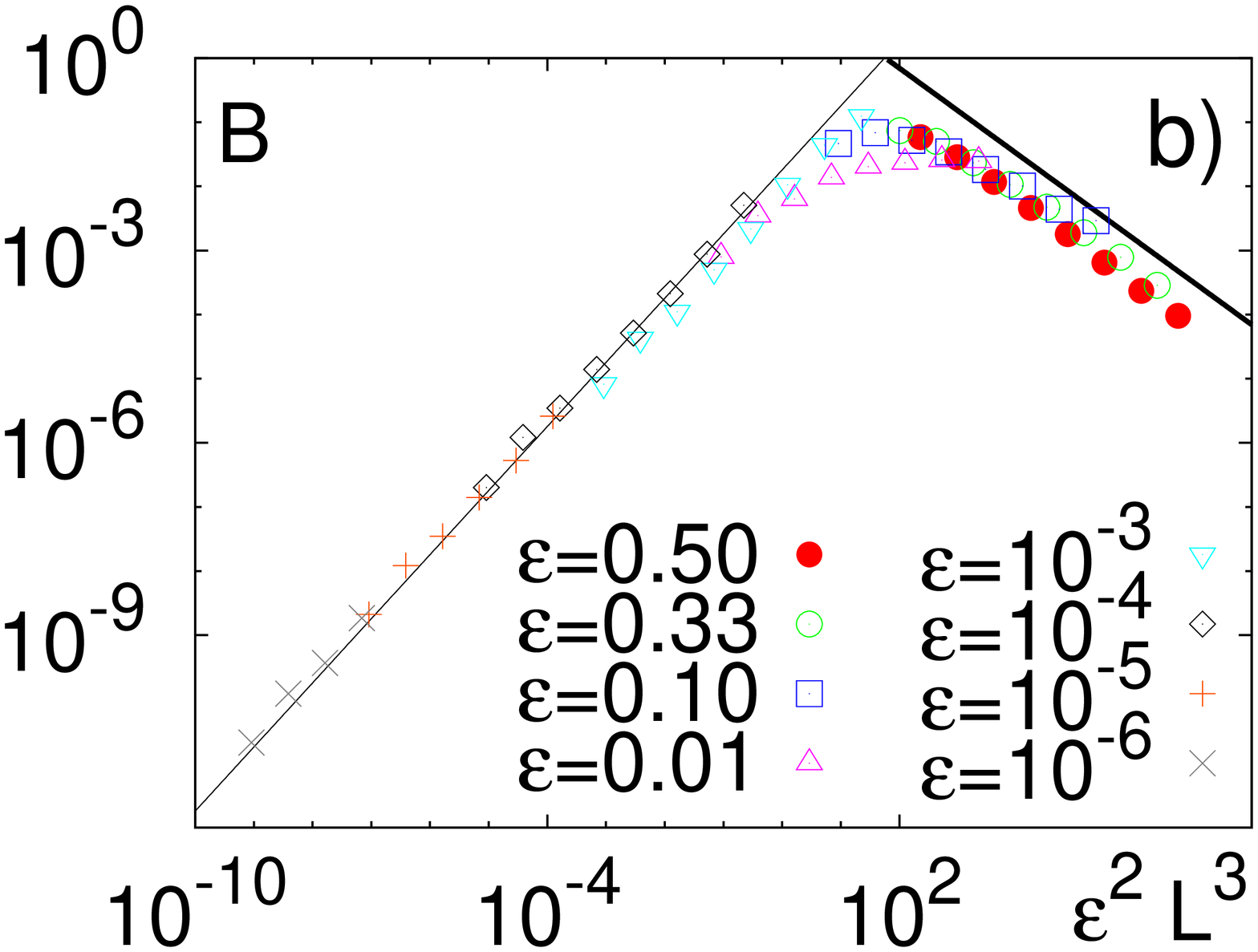,angle=270,width=0.5\lW} }
  \centerline{
    \psfig{file=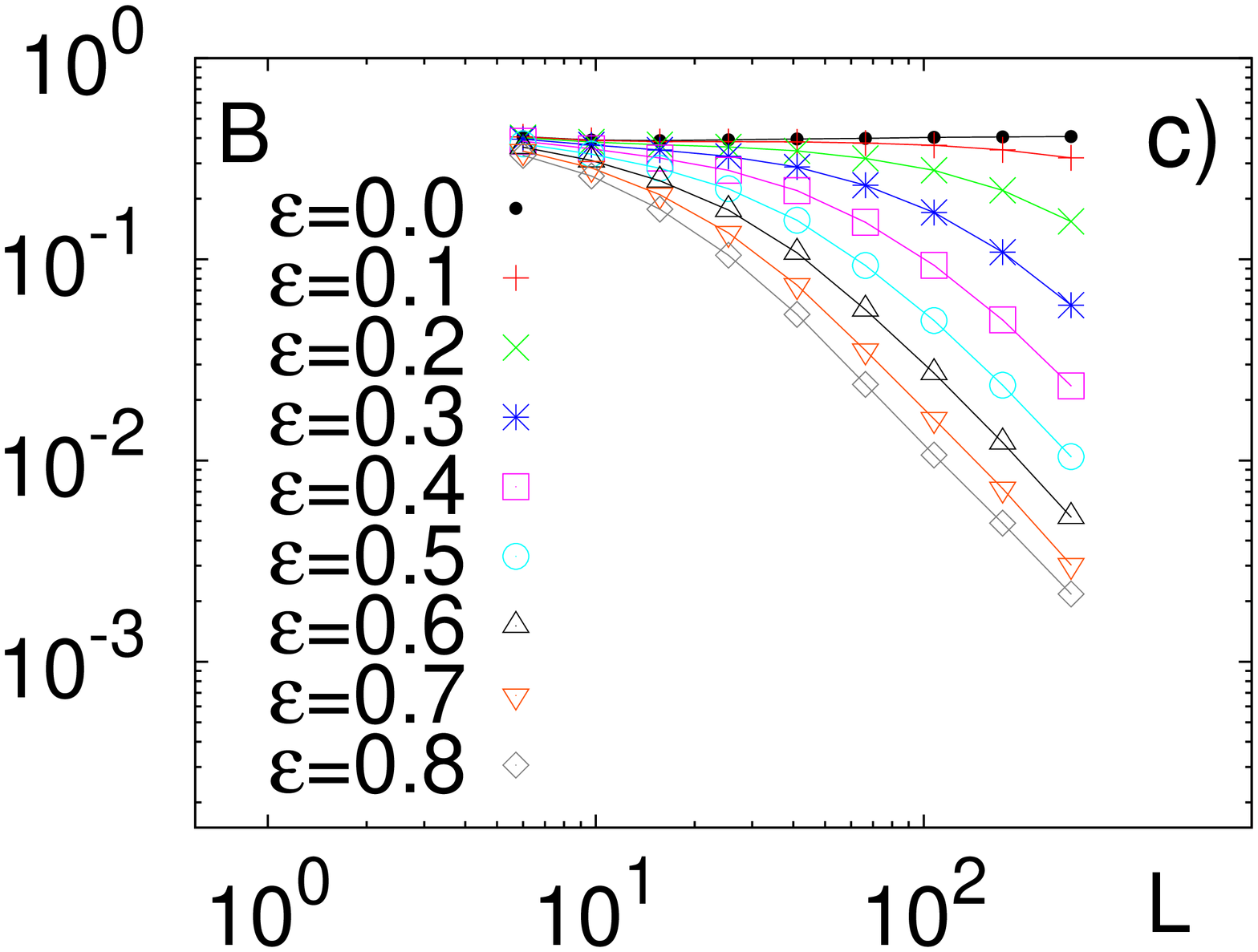,angle=270,width=0.5\lW}
    \psfig{file=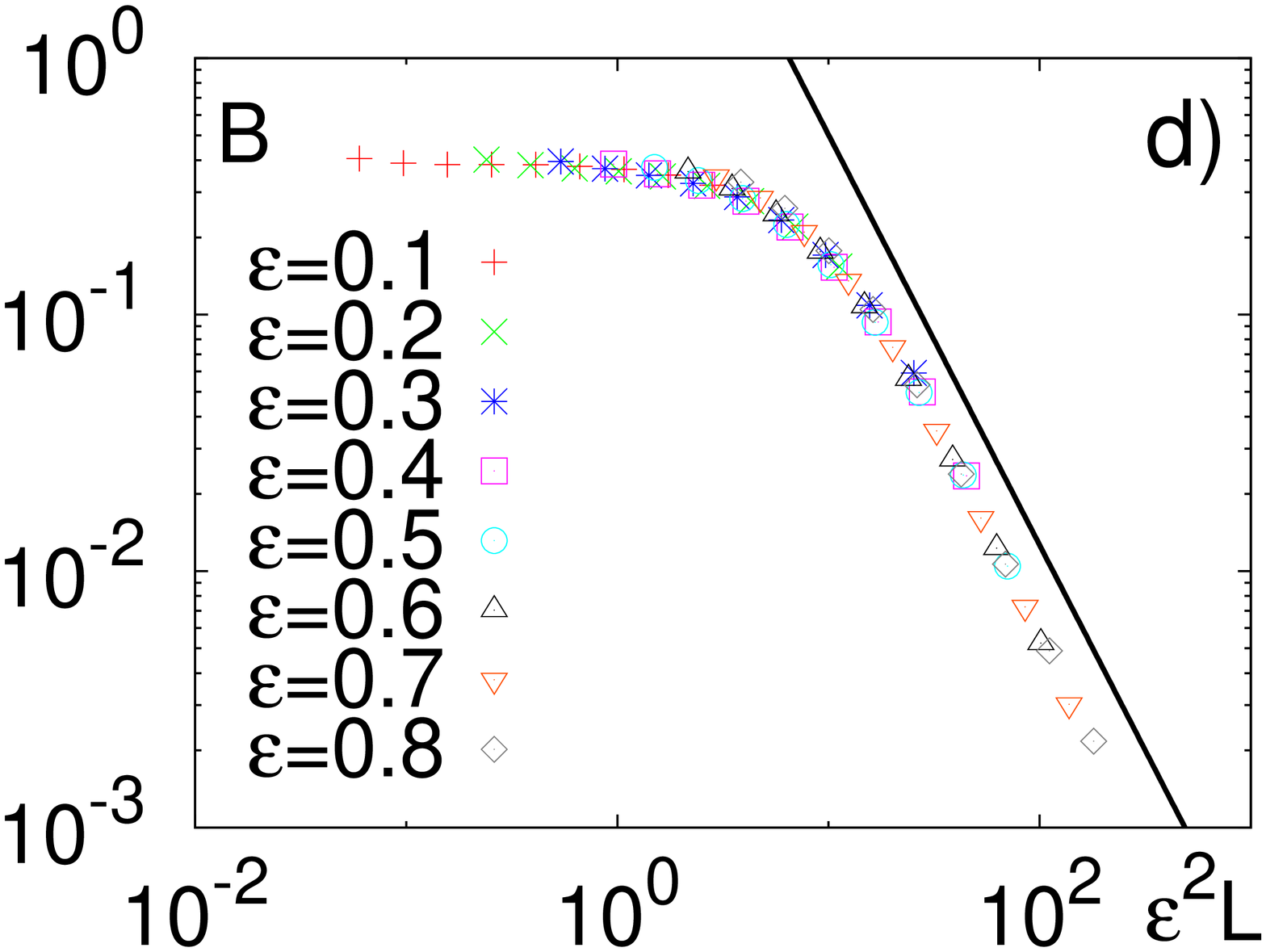,angle=270,width=0.5\lW}
  }
  \caption{Average bulk modulus $B(\epsilon,L)$ of randomly
    site-displaced Penrose approximants with period $L$ and disorder
    strength $\epsilon$, under Periodic (PBC, a) and b)) and Fixed
    (FBC, c) and d)) boundary conditions. Lines in a) and c) are
    guides to the eye. The thin line in b) is $B \sim \epsilon^2
    L^3$. Thick lines in b) and d) are, respectively, $B \sim
    (\epsilon^{2}L^{3})^{-2/3}$ and $B \sim (\epsilon^2 L)^{-1.6}$.}
\label{fig:1}
\end{figure}
The discovery of isostaticity in sphere packings~\cite{MIPT98} and
network glasses~\cite{TJCSIN00,VBTTR04} has inspired a great deal of
activity in the field of isostatic networks.  Recent
studies~\cite{MECI12,CFMEAI15} suggest that all elastic modulii of
geometrically disordered isostatic networks go to zero with increasing
linear size $L$, if disorder is uncorrelated.  Packings of hard
frictionless spheres or discs, on the other hand, have nonzero
compressive modulus $B$~\cite{OSLJAZ03,WOTR05}, despite being
isostatic~\cite{MIPT98} and disordered, because their contact network
is not random, but tuned to avoid negative forces.  Contact disorder
is correlated in these systems.  Attempts to model sphere packings as
randomly disordered isostatic networks have therefore failed.
However, in a recent letter~\cite{SLPTAJ15}, Stenull and Lubensky (SL)
claim that randomly disordered Penrose networks have nonzero $B$ for
large $L$. The present numerical study, using high-precision Conjugate
Gradient to solve the elastic equations shows their claim to be
incorrect, and clarifies the reason for their misinterpretation of the
data. \Figs{fig:1}a and b show $B(\epsilon,L)$ for Penrose periodic
approximants of orders 5 to 12 (up to $8 \times 10^4$ sites), whose
sites are randomly displaced within a circle of radius $\epsilon$. $B$
behaves roughly as $1/L^2$ for large $L$ (\Fig{fig:1}b).  However,
because \hbox{$B(\epsilon=0,L)=0\forall L$}~\cite{SLPTAJ15}, $B$ grows
as $\epsilon^2 L^3$~\cite{Note1} when \hbox{$\epsilon^2 L^3 <<
  10^2$}. The asymptotic regime $L>>L_0 \approx (10/\epsilon)^{2/3}$
is hard to reach for small $\epsilon$. This has been noted
already~\cite{CFMEAI15} for other disordered isostatic netwoks. The
data reported by SL~\cite{SLPTAJ15} (derived from normal-mode
calculations for a single, unspecified, value of $\epsilon$) are
similar to our results for $\epsilon=10^{-2}$ in \Fig{fig:1}a,
i.e.~$B$ \emph{appears to saturate}. Our scaling analysis in
\Fig{fig:1}b shows that this is a finite-size effect: the true
asymptotic behavior $B \sim L^{-2}$ would only be seen at much larger
sizes for this value of $\epsilon$.  Further validation of our claim
that $B \to 0$ for large $L$ is provided by the following: fixing a
line and a row of sites produces Penrose networks with Fixed
BC. $B^{\tiny FBC}$ is seen to go to zero with size when
\hbox{$\epsilon^2 L >> 1$} (see \Figs{fig:1}c and d). But $B^{\tiny
  FBC}$ is a rigorous upper bound for $B^{\tiny PBC}$. Therefore,
\hbox{$B^{\tiny PBC} \to 0$} for large $L$ as well. We additionally
mention that the effects of geometric disorder on elastic constants
can be described analytically for small $\epsilon$, giving rise to
rational expressions for $B(\epsilon,L)$, that predict an asymptotic
power-law behavior $B(\epsilon,L) \sim L^{-\mu}$ when $\epsilon \neq
0$. Details will be provided somewhere else~\footnote{C.~Moukarzel,
  2015, to be published.}.  We conclude that the main point raised by
SL~\cite{SLPTAJ15} is not justified: generic Penrose networks with
uncorrelated geometric disorder have zero bulk modulus for large
sizes. They are, therefore, no better suited to model jammed packings
than any of the previously studied isostatic networks with geometric
disorder.  The authors thank CGSTIC of CINVESTAV for computer time on
cluster Xiuhcoatl.
\end{document}